\documentclass{aa}  

\usepackage{graphicx}
\usepackage{txfonts}
\usepackage{natbib}
\bibpunct{(}{)}{;}{a}{}{,}

\newcommand{\doce}{\mbox{$^{12}$CO}}

\newcommand{\trece}{\mbox{$^{13}$CO}}

\newcommand{\jdu}{\mbox{$J$=2$-$1}}
\newcommand{\juc}{\mbox{$J$=1$-$0}}

\newcommand{\kms}{\mbox{km\,s$^{-1}$}}

\newcommand{\ms}{\mbox{$M_{\mbox{\sun}}$}}
\newcommand{\ls}{\mbox{$L_{\mbox{\sun}}$}}

\newcommand{\lsim}{\raisebox{-.4ex}{$\stackrel{\sf <}{\scriptstyle\sf \sim}$}}

\newcommand{\farcss}{\mbox{\rlap{.}$''$}}
\newcommand{\secp}{\mbox{\rlap{.}$''$}}
\begin{document}

   \title{Detection of Keplerian dynamics in a disk around the
     post-AGB star AC Her} 

   \author{V. Bujarrabal
          \inst{1}
          \and
          A. Castro-Carrizo\inst{2}
          \and
          J. Alcolea\inst{3}
\and
H. Van Winckel\inst{4}
}

   \institute{             Observatorio Astron\'omico Nacional (OAN-IGN),
              Apartado 112, E-28803 Alcal\'a de Henares, Spain\\
              \email{v.bujarrabal@oan.es} 
        \and
 Institut de Radioastronomie Millim\'etrique, 300 rue de la Piscine,
 38406, Saint Martin d'H\`eres, France 
        \and
             Observatorio Astron\'omico Nacional (OAN-IGN),
             C/ Alfonso XII, 3, E-28014 Madrid, Spain
         \and
Instituut voor Sterrenkunde, K.U.Leuven, Celestijnenlaan 200B, 3001
Leuven, Belgium
  }

   \date{Received 26 Jan 2015; accepted 04 Feb 2015}

 
  \abstract
   {}
{So far, only one rotating disk has been clearly identified and studied
  in AGB or post-AGB objects (in the Red Rectangle), by means of
  observations with high spectral and spatial resolution. However, disks
  are thought to play a key role in the late stellar evolution and
  are suspected to surround  many evolved stars. We aim to
  extend our knowledge on these structures.}
   {We present interferometric observations of \doce\ \jdu\ emission
     from the nebula surrounding the post-AGB star AC Her, a source
     belonging to a class of objects that share properties with the Red
     Rectangle and show hints of Keplerian disks.}
 {We clearly detect the Keplerian dynamics of a second disk orbiting an
  evolved star. Its main properties (size, temperature, central mass)
  are derived from direct interpretation of the data and model fitting.
  With this we confirm that there are disks orbiting the
  stars of this relatively wide class of post-AGB objects.}
   {}


   \maketitle
%

\section{Introduction}

Disks orbiting AGB and post-AGB stars are thought to play a major role
in the late AGB evolution and planetary nebulae formation. In the vast
majority of cases, the circumstellar envelopes around AGB stars are
roughly spherical and in isotropic expansion at moderate velocities (10
-- 20 \kms), see, for example,\ \cite{acc10}. Most protoplanetary and
planetary nebulae (PPNe, PNe) show a clear axial symmetry and fast
bipolar outflows ($\sim$ 100 \kms), which carry a good fraction of the
total mass ($\sim$ 0.1 \ms) and very large amounts of linear momentum,
see, for instance,\ \cite{bujetal01} and \cite{alc01}. This spectacular
transition takes place in a very short time, a thousand or a few
thousand years. The most widely accepted scenario to theoretically
explain this evolutionary phase implies that material is reaccreted by
the star or a companion from a rotating disk, followed by the launching
of very fast jets, in a process similar to that at work in forming
stars, see \cite{soker02}, \cite{blackman14}, and references therein.

Such orbiting disks have been frequently searched for, but their
detection has been extremely difficult. Except for one source
(the Red Rectangle, see below), in all AGB shells or post-AGB nebulae
in which spectroscopical observations allow measuring the velocity
field, only expansion has been detected. Many of them show oblate, even
very flat structures, but without any sign of rotation in the
observations. This is the case of axisymmetric shells around
semiregular variables \citep[e.g.,][]{libert10,acc10} and of flat
equatorial structures that are often found in bipolar PPNe
\citep[e.g.][]{alc07, acc02, acc12}.

Various indications of the existence of Keplerian disks have been
proposed. The very low expansion velocity (\lsim\ 1 \kms) sometimes
found in the equatorial structures mentioned above suggests that the
gas is ejected from a stable, and therefore probably Keplerian,
component at long distances to the star, in which outward
forces are weak.  On the other hand, a class of optically bright
post-AGB stars are found to show a remarkable NIR excess, which has
been attributed to the emission of hot dust at about 1000 K. The high
temperature would show that dust grains are kept close to the star,
again in a stable disk structure, see \cite{winckel03} and
\cite{aarle11}. These stars are found to be
systematically multiple \citep{winckel09} and the grains show a high
degree of evolution \citep[i.e.,\ a high critallinity;][]{gielen11},
which supports this interpretation.

   \begin{figure*}
   \centering{\resizebox{17cm}{!}{
   \includegraphics{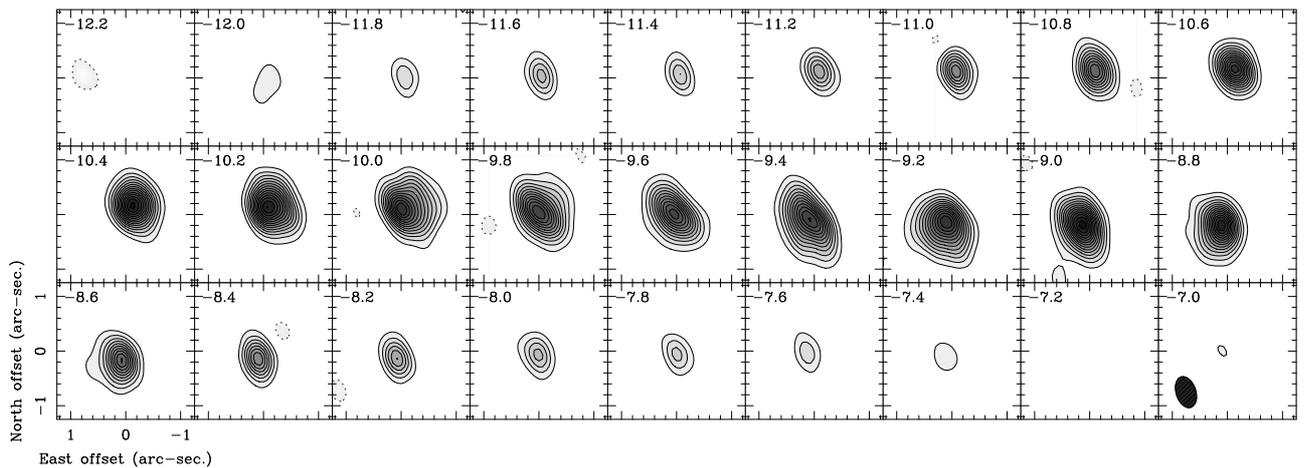}}}
   \caption{PdBI maps per velocity channel of \doce\ \jdu\ emission from AC
     Her. The $LSR$ velocity is indicated in the upper left
     corners. The first contour and  step are 16 mJy, negative
     contours are indicated by dashed lines. The dark ellipse in the 
     last panel shows the half-power extent of the synthetic beam.
The      conversion factor from Jy 
to K is 110 K per Jy/beam, the $rms$ noise measured in the maps is
about 4 mJy/beam.} 
              \label{maps}%
    \end{figure*}


Systematic observations of the \doce\ and \trece\ \jdu\ and \juc\ lines
in objects
with such a NIR excess \citep{bujetal13a} have shown narrow
line profiles formed by a single or double peak and 
weaker wings, which are very similar to those found in the
Red Rectangle. These profiles are also very similar to those in young
stars known to be surrounded by remnants of IS material in rotation and
to predictions of models of CO emission from orbiting disks, see
\cite{gui98}, \cite{gui13}, \cite{bujalc13}, etc. These results led
\cite{bujetal13a} to conclude that the CO lines in these sources
probably come from Keplerian disks.

However, the unambiguous detection of the Keplerian dynamics requires
very accurate spectroscopic observations with high spatial and spectral
resolution, able to describe the velocity field at subarcsec scales and
with spectral resolutions better than 0.5 \kms. To date, this type of
observations has resulted in the detection of rotation in just one
evolved object, the Red Rectangle, a well known PPN showing the above-mentioned NIR excess \citep{bujetal05,
  bujetal13b}. Radiointerferometric observations of CO rotational lines
undoubtedly showed a rotating equatorial disk, together with gas in
slow expansion that seems to be extracted from the disk. The main
properties of the nebula (extent, structure, dynamics) and gas physical
conditions (temperature, density) could be studied.

Several attempts to observe similar objects in this way have been
carried out \citep[particularly deep in 89 Her and
  IRAS\,19125+0343;][]{bujetal07}, but only expanding gas was
found. The detection of additional cases of rotating disks around
post-AGB stars is necessary to confirm our interpretation of
low-resolution CO data and that Keplerian disks are relatively common
among evolved nebulae. It is also necessary to study the main
properties of disks in a wide class of sources.

We present here the second firm detection of a Keplerian disk around a
post-AGB star, AC Her. AC Her, a binary star showing RV Tau
variability, is placed at a distance $D$ $\sim$ 1.6 kpc, with a total
luminosity $\sim$ 2500 \ls\ (Hillen et al.\ 2015). Its CO profiles
are narrow \citep{bujetal13a} and were attributed to rotating
structures. From optical and interferometric IR data, Hillen et
al.\ deduced the presence of a dust disk around AC Her with an inclination
with respect to the line of sight of about 45$^\circ$.

\section{Observations}

Observations of \doce\ \jdu\ line emission (230.538 GHz) in AC Her were
performed with the Plateau de Bure interferometer (project XB74). Data
were obtained with the 6Bq and 6Aq array configurations in March
2014. After a first calibration based on observations of nearby
quasars, data were self-calibrated by using the central continuum
source. A total flux of 30 mJy was obtained for the unresolved
continuum by averaging a 3.6 GHz bandwidth free of line emission in
each receiver polarization. A synthetic beam of 
0\secp 58$\times$0\secp 35 
in size (at half power, P.A.\ = 17$^\circ$) was obtained with data
natural weights, slightly worse than expected because of limited
$uv$-coverage in the extended baselines.

The CO results are presented in Fig.\ 1 with a channel spacing of 0.2
\kms, but the detailed data inspection was performed with spectral
resolutions of 0.1 \kms.  Note that the continuum emission was
subtracted from the data shown in Fig.\ 1 to better identify the weak
emission at extreme velocities.  From comparison with single-dish
profiles \citep{bujetal13a}, we deduce that 30--40\% of the flux was
filtered out in the interferometric maps. We hence detect the main and
most compact component in CO, which seems to be a disk whose projection
is elongated in the direction P.A.\ = 135$^\circ$. The
position-velocity diagram along this direction is shown in Fig.\ 2,
where the typical Keplerian rotation pattern is conspicuous, while the
plot in the perpendicular direction is completely symmetric with
respect to both velocity and position, as also expected for a Keplerian
disk.

\section{Results}

Some results can be directly and reliably extracted from our maps. It
is necessary to deduce these robust results independently of our model
fitting (Sect.\ 3.1), because, as we show below, such models contain
many uncertain parameters.

The velocity-position diagram along P.A.\ $\sim$ 135$^\circ$ is very
accurately coincident with those characteristic of Keplerian dynamics
\citep[for young and evolved objects, see,
  e.g.,][]{gui98,isella07,bujetal05,bujetal08}. This value gives the
inclination angle of the disk in the plane of the sky and cannot vary
by more than $\pm$ 5$^\circ$, otherwise the rotation features start to
appear also in cuts along the perpendicular direction.  We conclude
solely from inspecting this diagram that the detected CO emission from
AC Her comes from an orbiting disk. The non-negligible amount of flux
filtered out by the interferometer (Sect.\ 2) must come from a
relatively extended component, which could be gas in expansion, as
found in the Red Rectangle and 89 Her (Sect.\ 1).

If we assume that the \doce\ \jdu\ lines are opaque, which is
reasonable in view of the low \doce/\trece\ line ratio found in this
object \citep[][]{bujetal13a}, we can directly estimate from the
measured brightness the line excitation temperature. The values so
derived are a good estimate of the kinetic temperature, since low-$J$
lines are easily thermalized for the expected high densities \citep[see,
  e.g.,][and Sect.\ 3.1]{bujalc13}. The peak brightness in our maps is
$\sim$ 25 K, corresponding to a kinetic temperature $\sim$ 30 K
(after taking into account the conversion from brightness temperature
to Rayleigh-Jeans equivalent temperature for the \jdu\ frequency). This
is an average value within the synthetic beam; dilution can be
important in the inner (hot and rapidly rotating) regions, so this
value is probably more representative of the temperature in the outer
disk.  \cite{bujetal13a} assumed a typical temperature in the disks of
70 -- 100 K from data on the Red Rectangle; clearly, the disk around AC
Her is significantly cooler in spite of its smaller size (see below).

The measured extent of the emitting region, $\sim$ 1\farcss 5, is
significantly larger than that deduced from single-dish observations
\citep[0\farcss 7][]{bujetal13a} comparing single-dish
profiles with expectations for a disk with a temperature similar to
that of the Red Rectangle. This discrepancy is explained by the lower
temperature we find in AC Her. The disk around AC Her is much
smaller than that found in the Red Rectangle (about 6$''$ in diameter), however;
in linear units, the AC Her disk is about 2.5 times smaller (the Red
Rectangle is placed at $\sim$ 700 pc).

The total velocity range and separation between the brightness peaks in
AC Her are much smaller than in the Red Rectangle in spite of the
smaller size of the disk of AC Her. We can graphically see this effect
by comparing our Fig.\ 2 with Fig.\ 3 in \cite[][particularly the
  \juc\ line of the Red Rectangle, because of its lower size/resolution
  ratio]{bujetal05}. Provided that the disk around AC Her is not seen
face-on (Sect.\ 1), we deduce that the mass of the stellar component
must be lower than for the Red Rectangle; see model calculations in
Sect.\ 3.1.  No direct indication of expanding gas (in addition to the
Keplerian disk), as that found in the Red Rectangle and 89 Her, is seen
in our data. There is a weak absorption at about $-$11.5 \kms\ $LSR$
that could be due to gas in expansion, but its identification and
interpretation in these terms are very tentative. We hope that more
accurate observations will throw light on this question.

   \begin{figure}
     \hspace{.5cm}
{\resizebox{7cm}{!}{
   \includegraphics{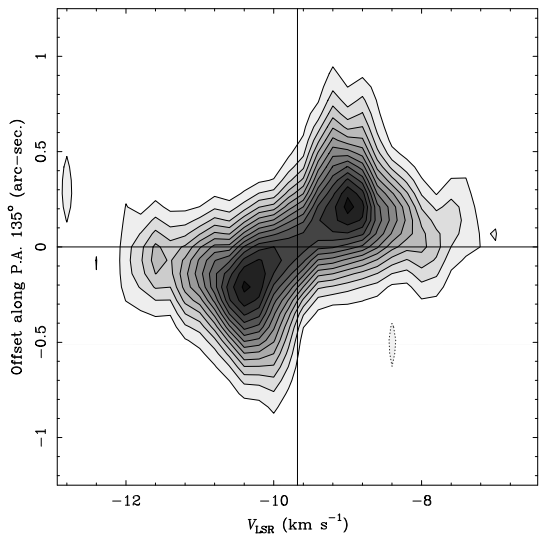}}}
   \caption{Position velocity diagram along the disk direction, P.A.\ =
     135$^\circ$. Color scale and contours are the same as in Fig.\ 1.}
              \label{pv}%
    \end{figure}

   \begin{figure}
     \hspace{.5cm}
{\resizebox{7cm}{!}{
   \includegraphics{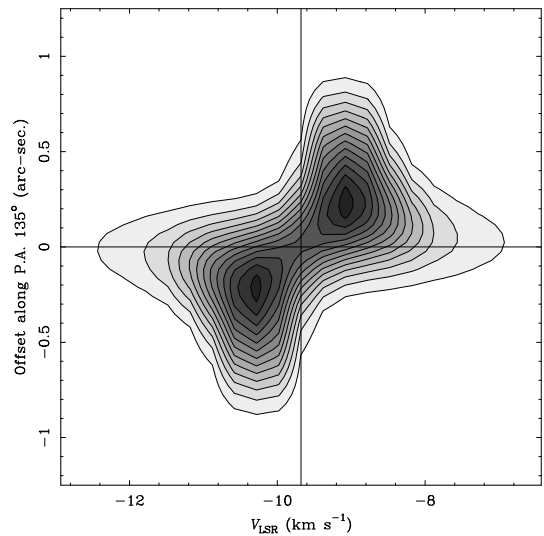}}}
   \caption{Position velocity diagram predicted by our best-fit
     model. All units, scales, and contours are the same as in Fig.\ 2.}
              \label{pv}%
    \end{figure}

\subsection{Model calculations}

Our modeling of the disk orbiting AC Her is based on that we performed
to explain our maps of the Red Rectangle. We used two codes. The first
one solves the 'exact' radiative transfer equations in two dimensions
for very many transitions and the level populations for very many
representative points in the disk.  A disk model is assumed, including
the disk structure and kinematics, as well as the distribution of the
main physical parameters. See the description of the calculations and
the detailed discussion of their accuracy in \cite{bujalc13}.
(However, our calculations show that the low-$J$ CO transitions are
thermalized for typical post-AGB disk conditions, as also concluded in
previous works; therefore, simpler excitation calculations, even
assuming LTE level populations, yield similar results.)  In a second
step, we solved the radiative transfer equation for the lines of sight
pointing at the observer, computed the predicted brightness
distribution in the plane of the sky and convolved it with the
interferometer synthetic beam. This led to maps and P-V cuts that can
be directly compared with those observed. A similar code was used in
previous papers \citep{bujetal05,bujetal08,bujetal13b}.

There are fewer data for AC Her than for the Red
Rectangle. Accordingly, we kept the model as simple as possible. We
took as starting point the disk structure found in the Red Rectangle
disk, but scaled it down by about a factor two since the disk around AC
Her is clearly smaller. We assumed no expansion, since no indication of
a departure from Keplerian dynamics has been found in our maps. We
assumed similar laws for the density and temperature, $n$ $\propto$
1/$r^2$ and $T$ $\propto$ 1/$r^{0.7}$, where $r$ is the distance to the
star, the proportionality constants being free parameters. The
inclination of the disk (with respect to line of sight) is a free
parameter, but it is assumed to range between 30 and 60 degrees
(Sect.\ 1). We show below that all these assumptions are confirmed to
be reasonable by our fitting, which leads to very satisfactory results
in spite of the few free parameters.

The results of our best fitting are summarized in Fig. 3. The agreement
with the observations is remarkably good and clearly confirms the
Keplerian dynamics of the disk around AC Her.

Our best-fitting model is described by $n$ = 3.5\,10$^5$ \,$(r_0/r)^2$
cm$^{-3}$, $T$ = 55\,$(r_0/r)^{0.7}$ K, $V_{\rm rot}$ = 1.6\,$\sqrt{r_0/r}$
\kms, and outer radius $R_{\rm out}$ = 1.7\,10$^{16}$ cm.  $r_0$ =
4.5\,10$^{15}$ cm, $r_0$ is the radius that defines the central decrease
of the disk thickness (see description in our previous works) and is
taken to be equal to the disk thickness (to further limit the free
parameters). We took the \doce\ abundance to be $X$(\doce) =
1.5\,10$^{-4}$, slightly smaller than that found for the Red Rectangle
because of the weak emission of the source and its low
\doce/\trece\ line ratio (despite the weak lines); this value is
practically an assumption because the low CO lines are
almost thermalized in our case and the density and the molecular
abundances are not independent parameters.

The inclination of the disk plane with respect to the north (P.A.) is
assumed to be 135$^\circ$, the disk position angle derived from our
observations.  We find a value of the disk inclination with
respect to the line of sight (which is equal to the inclination of the
disk axis with respect to the plane of the sky), $i$ $\sim$ 45$^\circ$,
compatible with indications from other studies (Sect.\ 1). Our
calculations indicate an uncertainty in this value of $\sim$
10$^\circ$ from comparing the predicted and observed extents in the
directions of the equator and of the rotation axis: for higher angles,
the predicted brightness extent in the axial direction is equal
to or
higher than that in the equator direction, in contrast to the
observations (we recall that the angular resolution is better in the
projected equator direction), while for low values of $i$ the contrast
is significantly higher than observed. The sense of $i$ is not given by
our data, nor do we  have other information on it.

From the disk dynamics and inclination derived here, we deduce that the
mass of the central stellar system is $\sim$ 1 \ms. This value is
somewhat lower than that derived from the binary movements in AC Her
\citep[$\sim$ 1.8 \ms,][]{winckel98} and than the value found from the
CO maps of the Red Rectangle (1.7 \ms). The difference can be due to
the relatively uncertain distance value, $D$, or to a slightly higher
inclination, $i$, within the acceptable range mentioned above. 
[Our central mass estimate is roughly proportional to $D$ and
cos$^{-2}$($i$).]

The total mass of the nebula, derived from integrating the disk
densities in the model, is also low, $M_{\rm neb}$ $\sim$ 1.2
10$^{-3}$ \ms, similar to the mass deduced for other similar objects
\citep{bujetal13a,bujalc13}, but lower than the mass of the Red
Rectangle disk, $\sim$ 6\,10$^{-3}$ \ms. It is slightly higher than
that derived for AC Her from low-resolution profiles by
\cite{bujetal13a}, 8.4 10$^{-4}$ \ms. This discrepancy mostly comes
from the longer distance used here (the derived mass value varies
approximately with $D^2$); but we recall the percentage of flux
resolved out in our interferometric observations and that, as our data
show, the temperature is lower than assumed in our previous work. It
is difficult to estimate the effects on the derived mass value from
this optically thick line because of the missing flux, but we can expect a 
slight underestimate of the mass. We
conclude that for $D$ = 1.6 kpc, the circumstellar mass of
the detected nebula around AC Her must be in the range 10$^{-3}$
\ms\ \lsim\ $M_{\rm neb}$ \lsim\ 2 10$^{-3}$ \ms.
We note that recent studies of the IR emission of AC Her (Hillen et
al.\ 2015) suggest a very compact inner region,
within 200 AU from the center and very thin,
which contains a high dust mass, $\sim$ 2.5 10$^{-3}$ \ms\ (with perhaps a
low gas-to-dust ratio). Our analysis can neither confirm nor rule out this
component, which would be unresolved in the maps and very optically
thick in \doce\ \jdu\ (and given the uncertainties in the dust disk
properties and in our modeling, particularly in the temperature of the
central regions).  The physical conditions derived here are just
representative of the extended disk probed in our maps, while the IR
data identify a very compact dust structure.  We recall that AC Her has
not been mapped in the \trece\ (less opaque) lines, hampering a proper
analysis of the nebular mass from model fitting.

\section{Conclusions}

   \begin{enumerate}

      \item We presented PdBI maps of \doce\ \jdu\ emission from the nebula
surrounding the post-AGB star AC Her, in which we convincingly
identified a Keplerian disk (Sects.\ 2 and 3).
Orbiting disks are suspected to play a fundamental role in the
evolution of AGB and post-AGB objects (Sect.\ 1), but before this work,
such a structure had been clearly detected and studied in only one
source (the Red Rectangle).

      \item 
AC Her and the Red Rectangle belong to a relatively wide class of
post-AGB stars in which rotating disks are suspected to exist and to
represent most of the nebular mass (Sect.\ 1). These are double stars
whose surrounding nebulae systematically show peculiar CO line profiles
(in observations with low angular resolution), similar to those
expected from disks in rotation. Our maps support that Keplerian disks
are systematically present in these sources and that they are,
therefore, relatively common in nebulae around binary evolved stars.

      \item 
We derived the main properties of the disk from directly interpreting
the data and model fitting.  We find a total disk
radius of $\sim$ 1.7\,10$^{16}$ cm.  Densities of 10$^{6}$ -- 10$^{4}$
cm$^{-3}$ and temperatures of 80 -- 20 K are obtained. The velocity
field is found to be basically Keplerian; we do not find clear evidence
of gas in expansion (as found in other similar objects).

\item 
We find a disk mass $\sim$ 1.5 10$^{-3}$ \ms\ and a central mass $\sim$
1.5 \ms, after accounting for possible underestimations due to
uncertainties in the modeling (Sect.\ 3.1).

\item
Significant differences between disks surrounding stars of this class
are found. The disk in AC Her is much smaller and contains less mass
than that of the Red Rectangle. It is also significantly cooler,
even though the gas is closer to the center. Our data do not show any sign
of outflowing gas, which is conspicuous in the Red Rectangle and
dominating in the nebula around the similar object 89 Her (and, of
course, in most nebulae around AGB and post-AGB stars, see Sect.\ 1).

\item
However, the existing data on AC Her are still meager, therefore
our modeling is uncertain. A fraction of the total flux is lost in the
interferometric maps, which may come from an extended outflow, and the
resolution is poor compared with the disk size. On the other hand, the
lack of maps of optically thin lines prevents a deeper analysis of the
total disk mass. We hope that future mapping of \doce\ and
\trece\ lines will allow more accurate studies.

   \end{enumerate}

\begin{acknowledgements}
      This work is based on observations carried out with the IRAM
      Plateau de Bure Interferometer; IRAM is supported by INSU/CNRS
      (France), MPG (Germany) and IGN (Spain). 
\end{acknowledgements}


{}

\end{document}